\documentclass[12pt,preprint]{aastex}
\usepackage{emulateapj5}
\submitted{Accepted for publication in ApJ}
\usepackage{natbib}

\newcommand{\fxunits}{\mbox{erg cm$^{-2}$ s$^{-1}$}}
\newcommand{\lxunits}{\mbox{erg s$^{-1}$}}

\def\ergps{erg s$^{-1}$ }
\def\hpone{h$_{\rm 50}$ }
\def\hmone{h$_{\rm 50}^{-1}$ }
\def\hmtwo{h$_{\rm 50}^{-2}$ }

\begin{document}

\title{The 160 Square Degree {\em ROSAT\/} Survey: the Revised Catalog
of 201 Clusters with Spectroscopic Redshifts\altaffilmark{1}}

\author{C.R.\ Mullis\altaffilmark{2,3},
        B.R.\ McNamara\altaffilmark{4,5},
        H.\ Quintana\altaffilmark{6}, 
        A.\ Vikhlinin\altaffilmark{5},
        J.P.\ Henry\altaffilmark{3},
        I.M.\ Gioia\altaffilmark{7,3},
        A.\ Hornstrup\altaffilmark{8},
        W.\ Forman\altaffilmark{5}, and
        C.\ Jones\altaffilmark{5}}

\altaffiltext{1}{Based on observations obtained at the W.M.\ Keck
Observatory, jointly operated by the California Institute of
Technology, the University of California and the National Aeronautics
and Space Administration, at the University of Hawai`i 2.2\,m
telescope, and at the European Southern Observatory 3.6\,m on La Silla
(ESO Programs 60.A-0694, 62.O-0586, 64.O-0455).}

\altaffiltext{2}{European Southern Observatory, Headquarters,
                 Karl-Schwarzschild-Strasse 2, Garching bei M\"unchen D-85748, 
                 Germany, cmullis@eso.org}
\altaffiltext{3}{Institute for Astronomy, University of Hawai`i, 
		 2680 Woodlawn Drive, Honolulu, HI 96822, USA}
\altaffiltext{4}{Department of Physics and Astronomy, Ohio University, 
                 Athens, OH 45701, USA}
\altaffiltext{5}{Harvard-Smithsonian Center for Astrophysics, 
                 60 Garden Street, Cambridge, MA 02138, USA}
\altaffiltext{6}{Departamento de Astronomia y Astrofisica, 
                 Pontificia Universidad Catolica de Chile, Casilla 104, 
                 Santiago, 22, Chile}
\altaffiltext{7}{Istituto di Radioastronomia del CNR, via Gobetti 101, 
                 Bologna, I-40129, Italy}
\altaffiltext{8}{Danish Space Research Institute, Juliane Maries Vej 30, 
		 Copenhagen 0, DK-2100, Denmark}

\shorttitle{160 DEG$^{2}$ {\em ROSAT\/} SURVEY CLUSTER CATALOG} \shortauthors{MULLIS ET AL.}


\begin{abstract}

We present the revised catalog of galaxy clusters detected as extended
X-ray sources in the 160 Square Degree {\em ROSAT\/} Survey, including
spectroscopic redshifts and X-ray luminosities for 200 of the 201
members.  The median redshift is $z_{\rm median}=0.25$ and the median
X-ray luminosity is $L_{\rm X,median} = 4.2 \times 10^{43}$ \hmtwo
\lxunits~ (0.5--2.0 keV).  This is the largest high-redshift sample
of X-ray selected clusters published to date. There are 73 objects at
$z > 0.3$ and 22 objects at $z > 0.5$ drawn from a statistically
complete flux-limited survey with a median object flux of $1.4 \times
10^{-13}$ \fxunits.  We describe the optical follow-up of these
clusters with an emphasis on our spectroscopy which has yielded 155
cluster redshifts, 110 of which are presented here for the first time.
These measurements combined with 45 from the literature and other
sources provide near-complete spectroscopic coverage for our survey.
We discuss the final optical identifications for the extended X-ray
sources in the survey region and compare our results to similar X-ray
cluster searches.
\end{abstract}


%
\section{Introduction} 

According to the theory of hierarchical structure formation, clusters
of galaxies are amongst the largest and most recent systems to form in
the matrix of cosmic construction
\citep[e.g.,][]{Peebles1993,Peacock1999}.  Measurements based on
dynamical, gravitational lensing, and \mbox{X-ray} data independently
confirm the extreme magnitude of cluster masses ($10^{14}$--$10^{15}$
$M_{\sun}$); thus situating them as the largest virialized masses in
the Universe \citep[e.g.,][]{Smail1997b,Wu1998,Allen2001,Clowe2002}.
Observed phenomena including cluster-cluster mergers, shock fronts,
infalling sub-clusters, and non-spherical morphologies attest to the
on-going assembly of clusters.  (e.g., \citealt{Gioia1999};
\citealt*{Vikhlinin2001}; \citealt{Czoske2002};
\citealt{Markevitch2002}; \citealt{Rose2002}).

As galaxy clusters form in the deepest gravitational potentials,
presumably at the intersections of filaments in the ``cosmic web'' of
large-scale structure, they are excellent tracers of the matter
distribution in the Universe (e.g., \citealt*{Bond1996};
\citealt{Jenkins1998}; \citealt{Mullis2001a}; \citealt{Borgani2001}).
Moreover, clusters are important tools for constraining cosmological
parameters.  For example, determinations of the mass, \mbox{X-ray}
luminosity, and \mbox{X-ray} temperature functions can be used to
measure the matter density parameter ($\Omega_{M}$) and the amplitude
($\sigma_{8}$) of density fluctuations
\citep[e.g.,][]{Henry2000,Borgani2001b, Vikhlinin2003}.  Observations
of the Sunyaev-Zeldovich (SZ) effect combined with \mbox{X-ray}
imaging of clusters can be used to assess the Hubble parameter
($H_{0}$) via a technique notably independent of the distance ladder
(e.g., \citealt*{Sunyaev1972,Silk1978,Carlstrom2002};).  By no means
is this overview of systemic cluster applications complete; see the
reviews, and references therein, of \citet{Sarazin1988} for a
historical perspective on cluster \mbox{X-ray} emission,
\citet{Mulchaey2000} for the \mbox{X-ray} properties of galaxy groups,
and \citet*{Rosati2002b} for the current
understanding of evolutionary trends in \mbox{X-ray} clusters of
galaxies.

The significance of galaxy clusters and the strong interest to
discover and investigate these remarkable structures are demonstrated
by the sheer number of surveys undertaken in recent years.  In the
high-energy domain alone, there are at least sixteen independent
\mbox{X-ray}-selected cluster surveys spanning the {\em Einstein\/}
and {\em ROSAT\/} eras (see \mbox{Table \ref{tab:xraysurveys}}).
\mbox{X-ray} selection is currently the optimal procedure for building
cluster samples with minimum bias and maximum statistical
completeness.  Techniques currently under development that may offer a
competitive alternative include gravitational lensing and SZ
surveys. \citet{Rosati2002b} provide a detailed discussion of
\mbox{X-ray} survey strategies, and \citet{Postman2002} offers a
comprehensive comparison of the \mbox{X-ray}, optical/NIR, and SZ
approaches.

We briefly review the basic properties of cluster \mbox{X-ray}
emission.  On megaparsec scales dark matter halos collapse pulling in
gas and galaxies.  This gas comprises $\sim$85\% of the luminous
cluster mass, and is heated to several $10^7$\,K through adiabatic
compression and shock heating during cluster formation.  At these
extreme temperatures the gas is highly ionized, optically thin ($n \sim
10^{-3}$\, cm$^{-3}$), and primarily radiates via thermal
bremsstrahlung  with minor contributions from thermal line transitions
(e.g., 6.7 keV Fe K$\alpha$).

The \mbox{X-ray} advantages of cluster selection are impressive.
Clusters are very luminous ($10^{43}$--$10^{45}$ \lxunits) and
spatially {\em extended} \mbox{X-ray} sources, which are efficiently
detectable to high redshifts \citep[e.g., RX\,J0848.9+4452 at
\mbox{$z=1.26$,}][] {Rosati1999,Stanford2001}.  Since the \mbox{X-ray}
emission is proportional to the gas density squared, clusters have
relatively peaked surface-brightness profiles.  This characteristic
coupled with the low \mbox{X-ray} background nearly eliminates
projection effects and underpins the high statistical quality of
\mbox{X-ray} samples.  The cosmological utility is further enhanced by
the strong correlation of \mbox{X-ray} luminosity with cluster mass,
and the fact that \mbox{X-ray} surveys feature well-defined selection
functions.  The latter is critical for reliably transforming source
counts to volume-normalized diagnostics.

In this paper we present the spectroscopic redshift catalog for the
160 Square Degree {\em ROSAT\/} Cluster Survey (hereafter 160SD).  The
201 galaxy clusters of the 160SD survey represent the largest,
high-redshift \mbox{X-ray} selected sample published to date.  First
described by \citet[hereafter V98]{Vikhlinin1998}, the 160SD clusters
have been used to study the evolution of cluster \mbox{X-ray}
luminosities and radii \citep{Vikhlinin1998b}, to present evidence for
a new class of \mbox{X-ray} overluminous elliptical galaxies or
``fossil groups'' \citep{Vikhlinin1999b}, to analyze the correlation
of optical cluster richness with redshift and \mbox{X-ray} luminosity
\citep{McNamara2001}, and to discover gravitational lenses
\citep{Munoz2001,Hornstrup2003}.  {\em Chandra} observations of
high-redshift 160SD clusters have been used to make an accurate
determination of the evolution of the scaling relations between
\mbox{X-ray} luminosity, temperature, and gas mass
\citep{Vikhlinin2002}, and to derive cosmological constraints from the
evolution of the cluster baryon mass function \citep{Vikhlinin2003}.
The most recent results are reported by \citet{Mullis2003b} who find
significant evolution in the number density of clusters as a function
of redshift based on number counts and changes in the \mbox{X-ray}
luminosity functions.

In \S\,2 we review the design of the 160SD survey and describe the
optical follow-up observations with an emphasis on the new spectroscopy.
In \S\,3 we present the revised cluster catalog which features
essentially complete spectroscopic redshifts.  We describe the general
properties of our cluster sample and compare them to the results of
other X-ray surveys in \S\,4. We close with a brief summary in \S\,5.
Throughout this analysis we assume an Einstein-\mbox{de Sitter}
cosmological model with $H_{0} = 50$ \hpone km s$^{-1}$ Mpc$^{-1}$ and
$\Omega_{M}=1$ ($\Omega_{\Lambda} = 0$), and quote \mbox{X-ray} fluxes
and luminosities in the {\mbox 0.5--2.0 keV} energy band.

%
\section{The 160 Square Degree ROSAT Survey}

The study of high-redshift galaxy clusters was revolutionized by the
{\em ROSAT\/} \mbox{X-ray} satellite \citep{Truemper1993} which was
used to perform the first all-sky survey with an imaging \mbox{X-ray}
telescope and to carry out a program of deep pointed observations for
nearly a decade (1990--1999).  These data, the {\em ROSAT\/} All-Sky
Survey \citep[RASS,][]{Voges1992} and the archive of pointed
observations, are the basis of many \mbox{X-ray} selected surveys
which have identified more than 1000 clusters; survey acronyms and
principal references are given in \mbox{Table \ref{tab:xraysurveys}}.
See \citet{Rosati2002b} for an extensive review of the subject.

Applying different combinations of solid angle and \mbox{X-ray} flux
sensitivity, the {\em ROSAT\/} cluster surveys provide good coverage
of the \mbox{X-ray} luminosity-redshift parameter space.  For example,
BCS, CIZA, NORAS, and REFLEX adopted the wide ($\sim$$10^4$ deg$^2$)
and shallow ($f_{\rm X} \ga 3 \times 10^{-12}$ \fxunits) strategy
using the RASS data to locate relatively nearby clusters ($z \la
0.3$).  Conversely surveys such as BMW-HRI, RDCS, SSHARC, and WARPS
executed deep searches ($f_{\rm X} \ga$ 2--6 $\times 10^{-14}$
\fxunits) with more modest solid angles ($\sim$20--200 deg$^2$) to
identify higher redshift clusters.  Our 160SD survey falls into the
latter category.  In the following section we review the design of the
program and elaborate on the optical follow-up observations.  See V98
for a complete description of the 160SD survey methodology.

\subsection{X-ray Selection of Cluster Candidates}

Clusters of galaxies are rare occurrences in the \mbox{X-ray} sky.
Surveys such as the EMSS and NEP which have rigorously identified {\em
all\/} \mbox{X-ray} sources within their survey regions find only
$\sim$12--15\% of the \mbox{X-ray} emitters at high Galactic latitudes
are clusters
\citep{Stocke1991,Henry2001,Mullis2001b,Gioia2003}. However, except
for nearby, isolated elliptical galaxies, clusters are essentially the
only extended \mbox{X-ray} sources away from the Galactic Plane.  Thus
with \mbox{X-ray} imaging data of sufficient spatial resolution and
signal-to-noise, \mbox{X-ray} extent is an excellent means to
efficiently identify galaxy clusters.

The 160SD clusters were selected based on the serendipitous detection
of extended X-ray emission in 647 archival {\em ROSAT\/} PSPC
observations.  The search included all fields with Galactic latitude
\mbox{$|b| > 30$$^{\circ}$} and hydrogen column densities
\mbox{$n_{\rm H} < 6 \times 10^{20}$ cm$^{-2}$} but excluded
10$^{\circ}$ radius regions around the SMC and LMC\@.  PSPC
observations targeting known clusters of galaxies, nearby galaxies,
star clusters, and supernova remnants were not analyzed.  A total of
160 deg$^{2}$ were surveyed at high fluxes. At lower fluxes, the sky
coverage smoothly decreases due to the survey selection function (see
Table 5 in V98). The area drops to 80 deg$^{2}$ at the median survey
flux ($1.2 \times 10^{-13}$ \fxunits) and to 5 deg$^{2}$ at $3.7
\times 10^{-14}$ \fxunits~ (which is the effective lower flux limit of
the survey).

Operating on hard-band images (0.6--2.0 keV), a wavelet decomposition
algorithm was used to detect X-ray sources between 2.5\arcmin\, and
17.5\arcmin\, from the PSPC field center.  The inner radius allows for
the avoidance of the original target of the pointing and the outer
radius is set by the window support structure of {\em ROSAT\/} PSPC.
X-ray sources significantly broader than the point-spread function
(PSF), as determined through a maximum-likelihood determination,
qualified for the initial list of candidate galaxy clusters.  

The PSF
of the {\em ROSAT\/} PSPC has a FWHM of about 25\arcsec\, on-axis
which grows to about 45\arcsec\, at 17.5\arcmin\, off axis.
The radial surface-brightness profile of cluster X-ray emission is
characterized by a $\beta$-model,
\mbox{$I(r,r_c)=I_{0}(1+r^2/r_c^2)^{-3\beta + 0.5}$} \citep{Cavaliere1976}.
A cluster with the canonical properties (core radius $r_{c}$=250
\hmone kpc and slope $\beta=2/3$) at $z=1$ corresponds to an angular
radius of $\sim$30\arcsec\ or a FWHM of $\sim$60\arcsec\, and thus is
easily resolved in our survey.  For a more stringent scenario consider
$r_{c}$=100 \hmone kpc which is at the very low end of the observed
distribution of core radii \citep[e.g.,][]{Jones1999}.  At $z=1$ the
intrinsic radius is $\sim$12\arcsec.
Nonetheless the convolution of the
intrinsic profile and the PSF results in a significantly extended
source 
even for this compact cluster at high redshift.  The decreased
detection probability for such a small cluster at larger off-axis
angles is incorporated into the 160SD selection function.

It is important to recognize that though X-ray extent is the primary
selection criterion for our survey, comparisons with other surveys
\citep[e.g., WARPS,][]{Perlman2002} demonstrate that no clusters were
missed as unresolved sources.  Hence the results of the 160SD survey
are in effect a statistically complete, X-ray flux-limited cluster
sample.

\subsection{Initial Optical Follow-up Observations}

\begin{figure}
{\epsscale{0.5}
\plotone{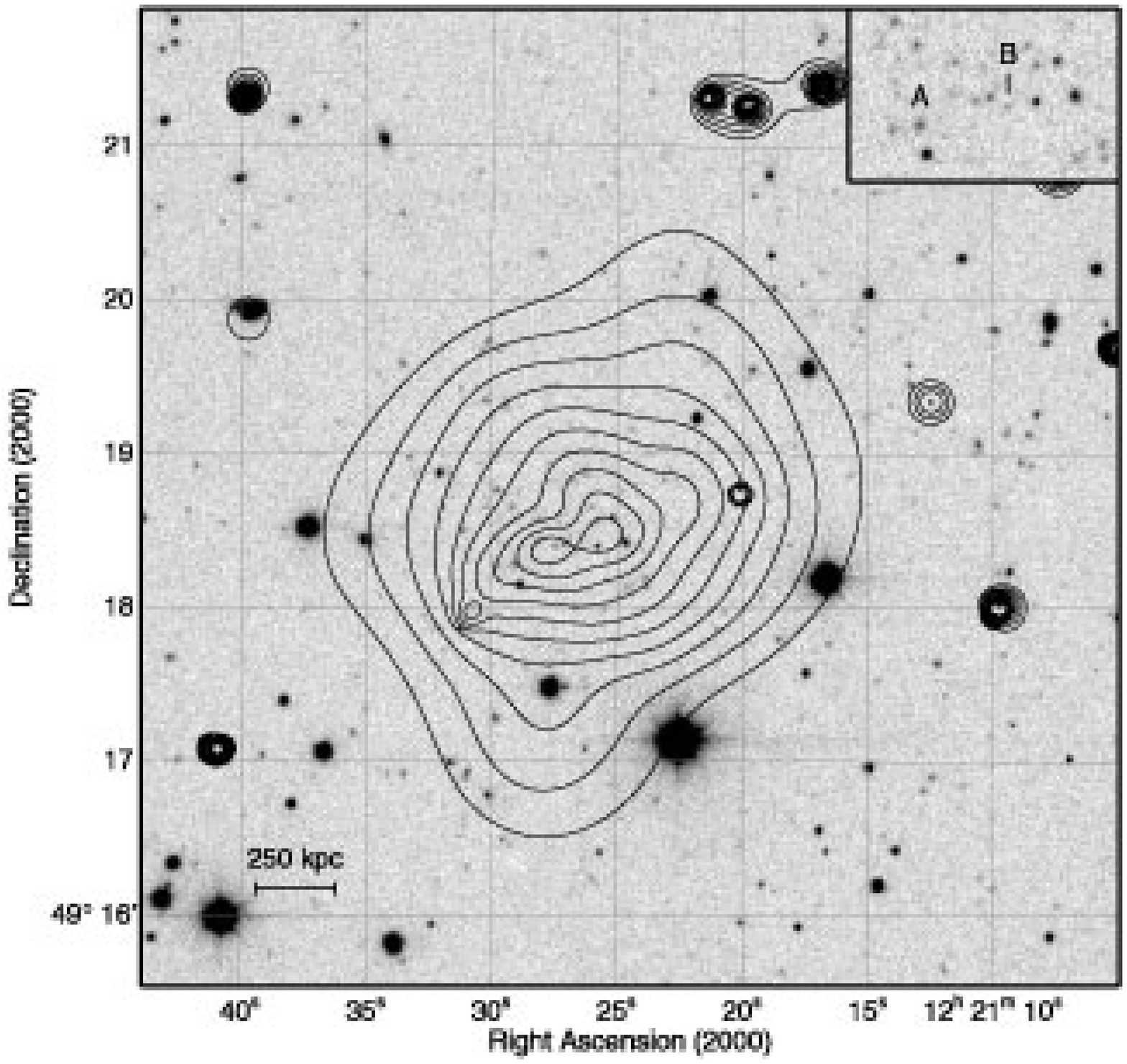}
\caption{RX\,J1221.4+4918 (\#119): a distant cluster at
$z=0.700$. A 5 minute $R$-band image taken with the FLWO 1.2m on
25 Jan 1998 is overlaid with adaptively smoothed X-ray flux contours
in the 0.7--2.0\,keV band from an 80\,ks observation with the {\em
Chandra} ACIS-I.  Contours are logarithmically spaced by factors of
1.4 with the lowest contour a factor of 2 above the background ($5.5 \times 10^{-4}$ counts s$^{-1}$ arcmin$^{2}$).
Features in the X-ray contours are significant at a level of $\ga 4
\sigma$.  The inset (upper-right) indicates the cluster galaxies for
which redshifts were measured (see spectra in \mbox{Figure
\ref{RXJ1221.4+4918_spectra}}). The $R$-band magnitude of galaxy A is
20.23. The scale bar shows the angular size of 250 kpc at
$z=0.700$.\label{RXJ1221.4+4918}}}
\end{figure}

\begin{figure}
\plotone{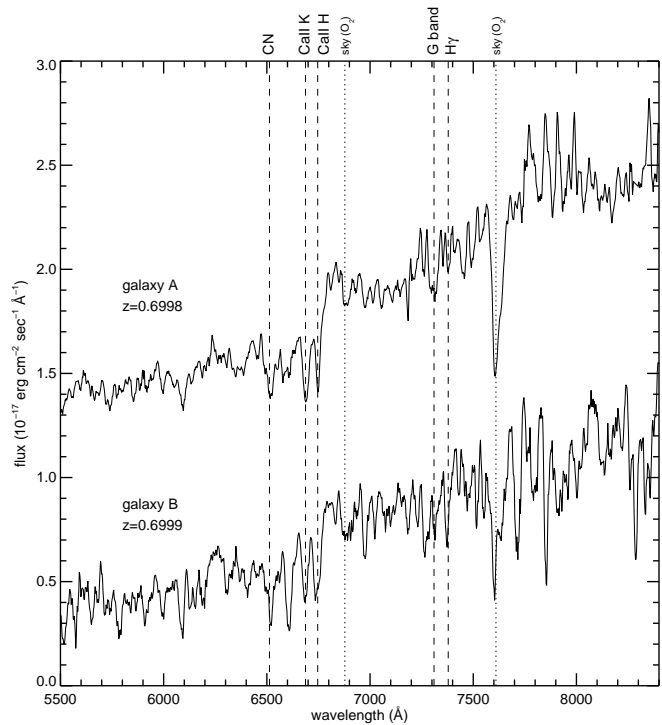}
\caption{RX\,J1221.4+4918 (\#119): a distant cluster at
$z=0.700$. Longslit spectra of galaxies A and B measured with Keck-II
LRIS on \mbox{08 Jul 1999}.  Total integration time was 30 minutes.  Galaxy
A is vertically displaced by $1\times 10^{-17}$ and galaxy B is scaled
by a factor of 2 for an optimal comparison of the spectra.  The measured
redshifts are $z_{\rm A}=0.6998 \pm 0.0004$ and $z_{\rm B}=0.6999
\pm 0.0008$.  The dashed lines indicate the positions of stellar
absorption features at the cluster redshift ($z=0.700$) and the dotted
lines mark the wavelengths of atmospheric absorption bands.
\label{RXJ1221.4+4918_spectra}}
\end{figure}

Following the X-ray selection of 223 candidate clusters, optical
observations are required to verify the presence of a galaxy cluster
and to estimate the distance to each system.  V98 described the
initial phase of these follow-up observations.  A search of the
NASA/IPAC Extragalactic Database
(NED\footnote{http://nedwww.ipac.caltech.edu/}) revealed 37 previously
known clusters, 31 of which had measured redshifts in the
literature. For the remaining 186 candidates, photometric CCD imaging
in the $R$-band was obtained using the Fred Lawrence Whipple
Observatory (FLWO) 1.2m, European Southern Observatory (ESO) / Danish
1.54m, and Las Campanas 1m telescopes.  These were supplemented in the
case of seven bright clusters using the DSS-II plates which can be
used to identify clusters to $z \la 0.45$.  The CCD imaging was
sufficiently deep to reveal clusters to $z = 0.7-0.9$.  \mbox{Figure
\ref{RXJ1221.4+4918}} illustrates the relative ease of establishing
the presence of distant clusters even with short exposures through a
modest telescope.  An overdensity of galaxies visible in this 5 minute
$R$-band image from the FLWO 1.2m is perfectly aligned with the X-ray
emission from an 80\,ks $Chandra$ observation.  Our subsequent
\mbox{Keck-II} longslit spectra (Figure \ref{RXJ1221.4+4918_spectra})
taken along the major axis of the X-rays and galaxy distribution
confirm the presence of a cluster at $z=0.700$.  If a concentration of
galaxies is not present in the CCD imaging it was classified as a
false detection though it could presumably be a very distant cluster
(e.g., RX\,J0848.9+4452 which is 160SD cluster \#61 at
\mbox{$z=1.26$}).

In total 203 clusters were confirmed by V98 (see \mbox{Table
\ref{tab:identifications}}). Of the remaining 20 sources, 19 are
probable false detections resulting from the blends of unresolved
X-ray point sources.  It was not possible to confirm
\mbox{RX\,J1415.6+1906 (\#157)} because it is obscured by the glare of
Arcturus.  The exceptional quality of the 160SD X-ray selection is
reinforced by this high success rate (91\%).

The second phase of the optical follow-up concerns spectroscopic
measurements principally used to measure distances to the clusters.  Of course the
determination of concordant redshifts from amongst the likely cluster
members further justifies the cluster classification, but this is
somewhat ancillary given the presence of extended X-ray emission
spatially coincident with a galaxy density enhancement.  As previously
noted, spectroscopic redshifts for 31 of the clusters were available
from the literature.  V98 obtained spectroscopic redshifts for 45
additional clusters using the Multiple Mirror Telescope 6$\times$1.8m,
ESO 3.6m, and ESO/Danish 1.54m telescopes, and presented photometric
redshifts for 124 of the remaining 127 clusters based on the magnitude
of the brightest cluster galaxy (BCG).  Reliable photometric estimates
were not possible for three candidates \mbox{RX\,J0910.6+4248 (\#69)},
\mbox{RX\,J1237.4+11141 (\#122)}, and \mbox{RX\,J1438.9+6423 (\#164)}
because the selection of the BCG in each case was unclear due to the
large angular extent of the candidate.  Note the object identification
numbers in parentheses are the same as the ones used in V98.  The distance estimates of V98 are summarized in the second column of \mbox{Table \ref{tab:redshifts}}.

The membership of the 160SD cluster sample has proven to be remarkably
stable since first presented by V98.  Based on the results of
extensive spectroscopic follow-up (described in \S\,\ref{newzs}), the
revised sample consists of 201 confirmed clusters, 21 false
detections, and one source obscured by Arcturus (see \mbox{Table
\ref{tab:identifications}}).  These classifications have changed very
little; one probable false detection has been identified as a cluster
and three cluster candidates have been reclassified as probable
falses.  RX\,J0848.9+4452 (\#61) was originally classified as a
probable false detection because no galaxy overdensity was visible in
our optical image of this field.  However, it was forewarned that such
an object could be a very distant ($z \ga 0.9$) cluster.  In fact
\citet{Rosati1999} have proven that this X-ray source is a cluster at
$z=1.26$ making it the most distant system found in any of the X-ray
surveys thus far.  \mbox{RX\,J0857.7+2747 (\#65)},
\mbox{RX\,J1429.6+4234 (\#163)}, and \mbox{RX\,J2004.8-5603 (\#197)}
were initially cluster candidates but we now interpret them to be
likely false detections because our spectroscopic survey of these
fields failed to find any coherent structure in redshift space.

\subsection{New Spectroscopic Redshifts}
\label{newzs}

Since the initial follow-up observations we have gone on to measure
spectroscopic redshifts for 110 additional clusters from our 160SD
survey using the Keck-II 10m and the University of Hawai`i (UH) 2.2m
telescopes at the Mauna Kea Observatories, and the ESO 3.6m telescope
at La Silla Observatory.  Combining these new redshifts with 76
measurements reported by V98 and 14 redshifts from the literature and
private communications results in essentially complete (200 of 201
clusters) spectroscopic coverage for our entire sample.  Refer to
\mbox{Table \ref{tab:redshifts}} for a summary of the redshift
determinations. The new cluster catalog will be presented in
\S\,\ref{updatedsample}.  Here we give a technical description of the
new observations.

We chose the targets for the low-dispersion spectroscopic observations
using our previously described CCD images.  This was usually a
straightforward task since in most cases unambiguous cluster members
were spatially coincident with the extended X-ray emission.  Longslit
spectra were obtained with the Keck-II 10m and the UH 2.2m resulting
in usually 2--3 concordant galaxy redshifts per cluster, and always
including the BCG.  Longslit and multi-object spectra were taken with
the \mbox{ESO 3.6m}, the latter producing 10--15 galaxy redshifts
per cluster.

\begin{figure}
\plotone{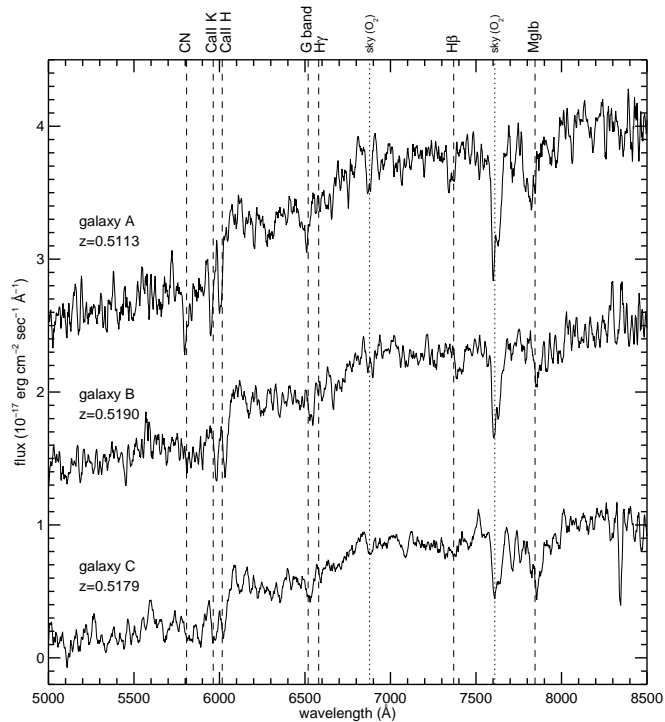}
\caption{RX\,J1524.6+0957 (\#170): a nearby group ($z=0.078$) and a
distant cluster ($z=0.516$). Longslit spectra of galaxies A, B, and C
measured with Keck-II LRIS on 27 Jan 2000.  Total integration time was
20 minutes.  Galaxies A and B are vertically displaced by $2\times
10^{-17}$ and $1\times 10^{-17}$, respectively, for an optimal
comparison of the spectra.  The measured redshifts are $z_{\rm A}=0.5113 \pm 0.0002$, $z_{\rm B}=0.5190 \pm 0.0005$ and $z_{\rm C}=0.5179 \pm 0.0004$.  The dashed lines indicate the positions of
stellar absorption features at the cluster redshift ($z=0.516$) and
the dotted lines mark the wavelengths of atmospheric absorption
bands.\label{RXJ1524.6+0957_spectra}}
\end{figure}

We measured redshifts for the most distant clusters ($z\ga0.5$) using
the Keck-II 10m telescope with the Low Resolution Imaging Spectrometer
\citep[LRIS, ][]{Oke1995} during four nights of observations (1999
July 07--08 and 2000 January 26--27). Examples are shown in Figures
\ref{RXJ1221.4+4918_spectra} and \ref{RXJ1524.6+0957_spectra}.  The
300 lines~mm$^{-1}$ grating (5000\AA\ blaze) combined with the GG495
long-pass order-blocking filter yielded an effective wavelength
coverage of approximately 5000\AA--9000\AA\, and a dispersion of
2.45\AA{ }pixel$^{-1}$.  With the Tektronix 2048$\times$2048 CCD
detector the spatial scale is 0.215\arcsec\, pixel$^{-1}$. We used a
slit of 1.5\arcsec\, in width, which gives a reduced spectral
resolution of $\sim$16\AA~FWHM ($R \sim 440$).  Exposure times were
typically 1200s--2400\,s and the seeing was \mbox{0.7\arcsec --
1.1\arcsec}.

Moderately distant clusters were observed using the \mbox{UH 2.2m}
with the Wide Field Grism Spectrograph (WFGS) during ten nights of
spectroscopy (1998 May -- 2000 February).  The instrument setup
consisted of the Tektronix 2048$\times$2048 CCD at the f/10 focus with
the 420 lines~mm$^{-1}$ red grism and a 1.8\arcsec\, slit, providing a
wavelength coverage of approximately {3800\AA--9000\AA}.  The spectral
dispersion is 3.6\AA{ }pixel$^{-1}$, the spectral resolution is
$\sim$19\AA~FWHM ($R \sim 340$), and the spatial scale is
0.35\arcsec\,pixel$^{-1}$.  The typical seeing was
0.8\arcsec--1.5\arcsec\, and integration times ranged between
600s and 3600\,s.

We observed the southern clusters not visible from Mauna Kea using the
\mbox{ESO 3.6m} with the ESO Faint Object Spectrograph and Camera
(EFOSC2), in both the longslit and the multi-object spectroscopy modes.
Data were taken over 6 nights (\mbox{1998 March 28,} \mbox{1999 February
15--16,} and \mbox{1999 November 06--08}) during which time the seeing
varied over the range 0.6\arcsec--1.8\arcsec.  The 300 lines~mm$^{-1}$
grating (\#6, 5000\AA\, blaze) with a 1.5\arcsec\ slit provides a
wavelength coverage of about 3900\AA--8000\AA\, with a spectral
resolution of $\sim$20\AA\,\mbox{($R \sim 300$)}.  The spectral
dispersion is 4\AA{ }pixel$^{-1}$ and the spatial scale is
0.31\arcsec\,pixel$^{-1}$ using the Loral 2048$\times$2048 CCD binned
2$\times$2 at readout.  Exposure times were usually between 900s and
1800s.

In total we observed over 400 individual galaxies to derive the 110
new cluster redshifts reported in this paper.  The only cluster for
which we lack a redshift is RX\,J1237.4+1141 (\#122).  The unusually
large extent of this source would require an extensive redshift survey
to yield a reliable distance measurement. We analyzed our data
following standard procedures using IRAF\footnote{IRAF is distributed
by the National Optical Astronomy Observatories, which are operated by
the Association of Universities for Research in Astronomy, Inc., under
cooperative agreement with the National Science
Foundation. http://iraf.noao.edu} reduction packages and
IDL\footnote{http://www.rsinc.com,
http://idlastro.gsfc.nasa.gov/homepage.html} routines.
Two-dimensional spectra were de-biased and flat-fielded.
One-dimensional spectra, with the sky background subtracted, were
extracted and wavelength calibrated.  Finally the instrumental
response was removed using observations of the spectrophotometric
standard stars of \citet{Oke1983} and \citet{Massey1988}.  Redshifts
were measured based on the offsets of absorption features commonly
observed in early-type galaxies including the Ca II H and K doublet
(3933.68\AA,{ }3968.49\AA), the 4000\AA~break, the CH G band
($\sim$4300\AA), Mg\,Ib ($\sim$5175\AA), and Na\,Id (5889.97\AA).

Brief descriptions of a few individual clusters demonstrate the
reliability of our X-ray source characterization and the rigor of our
optical follow-up program.  Though relatively simple X-ray/optical
configurations such as that shown in \mbox{Figure
\ref{RXJ1221.4+4918}} are the most commonly observed, there are more
challenging scenarios.  Take for instance the 160SD source
RX\,J0921.2+4528 (\#70).  We were surprised when our spectrum of the
apparent BCG revealed a broad-line QSO at $z=1.66$.  Since active
galactic nuclei (AGN) are usually X-ray luminous we could have stopped
here and classified RX\,J0921.2+4528 as an AGN.  However our analysis
of the X-ray data shows this source is significantly ($6\sigma$)
broader than the {\em ROSAT\/} PSPC PSF thus motivating us to pursue
this further.  With subsequent spectroscopy we found five concordant
redshifts at $z=0.315$ confirming the cluster identification.  More
interestingly, a second QSO at $z=1.66$ was found separated
6.93\arcsec\, from the first.  \citet{Munoz2001} further describe this
wide-separation gravitational lens candidate.  Our Monte-Carlo
simulations show that a 90\% upper limit on the flux of a point source
at the QSO position is about 35\% of the total cluster flux.

\begin{figure}
{\epsscale{0.5}
\plotone{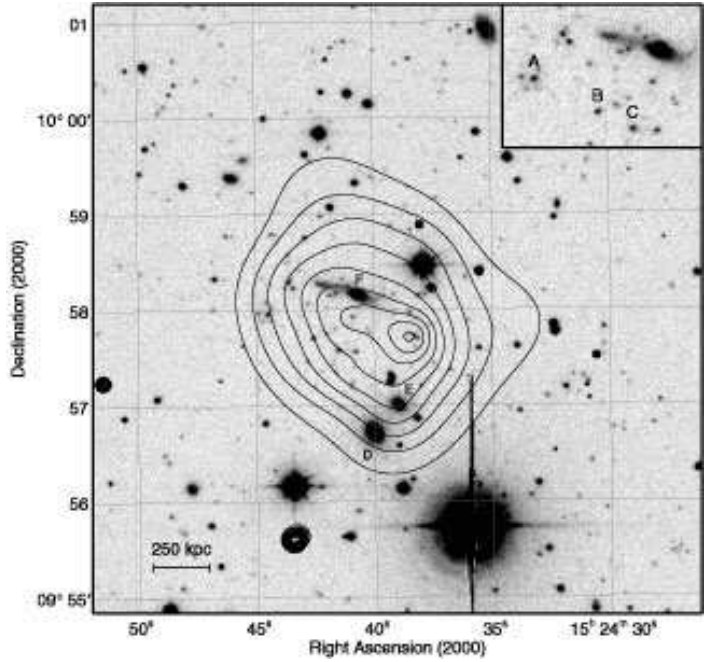}
\caption{RX\,J1524.6+0957 (\#170): a nearby group ($z=0.078$) and a
distant cluster ($z=0.516$).  A 10 minute $I$-band image taken
with the FLWO 1.2m on 08 Jul 1997 is overlaid with adaptively smoothed
X-ray flux contours in the 0.7--2.0\,keV band from a 51\,ks
observation with the {\em Chandra} ACIS-I.  Contours are
logarithmically spaced by factors of 1.4 with the lowest contour a
factor of 2 above the background ($9.7 \times 10^{-4}$ counts s$^{-1}$ arcmin$^{2}$).  Features in the X-ray contours are
significant at a level of $\ga 4 \sigma$.  The inset (upper-right)
indicates the galaxies (A--C) of the distant cluster for which
redshifts were measured (see spectra in \mbox{Figure
\ref{RXJ1524.6+0957_spectra}}). The foreground group galaxies (D--F) are at redshifts $z_{\rm D}=0.0787 \pm 0.0009$, $z_{\rm
E}=0.0767 \pm 0.0009$, and $z_{\rm F}=0.08$? (K. Romer, priv. comm.).
The $R$-band magnitude of galaxy C is 20.29. The scale bar shows the
angular size of 250 kpc at $z=0.516$.\label{RXJ1524.6+0957}}}
\end{figure}

Another complex source to disentangle is RX\,J1524.6+0957 (\#170).
With a cursory inspection of the $I\/$-band image shown in
\mbox{Figure \ref{RXJ1524.6+0957}} one might conclude the obvious
low-redshift group is the optical counterpart of the X-ray source.
This interpretation was reported by the BSHARC survey who quoted a
redshift of $z=0.078$ \citep{Romer2000}.  However, closer scrutiny of
the fainter objects suggests a concentration of galaxies at the X-ray
position.  Our Keck-II spectroscopy for these faint galaxies
establishes a distant cluster at $z=0.516$ (\mbox{Figure
\ref{RXJ1524.6+0957_spectra}}).  Were the nearby group the principal
source of the X-rays, it would have to be exceedingly compact.  The measured
core radius of 26\arcsec\, corresponds to 52 \hmone kpc at $z=0.078$
or 190 \hmone kpc at $z=0.516$.  The former is unrealistic whereas the
latter is quite typical.  Finally, we measure an \mbox{$\sim$5 keV}
temperature for the intracluster medium, via a spectral fit to the
{\em Chandra\/} data, which strongly rejects the group-dominant
scenario.  Thus we conclude the primary identification for
RX\,J1524.6+0957 is a distant cluster at $z=0.516$.

\begin{figure}
\plotone{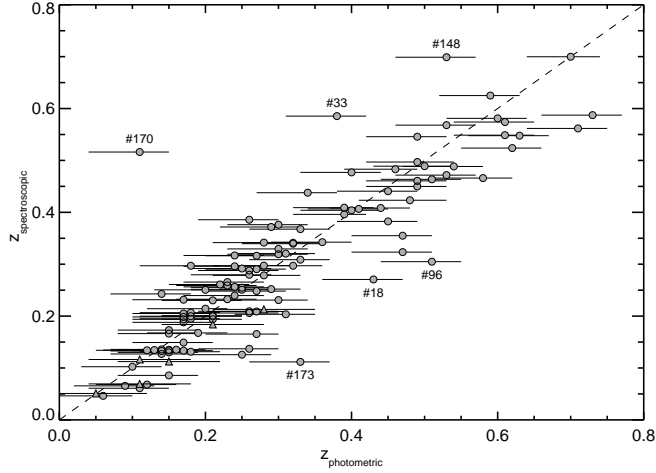}
\caption{Spectroscopic redshift versus photometric redshift for the
124 clusters of the 160SD sample originally published with photometric
redshifts in V98.  The circles represent the 117 objects for which the
photometric redshift estimates were based on CCD imaging; whereas the
triangles represent the 7 objects for which photometric redshifts were
based on DSS images.  The horizontal error bars are the 90\%
confidence intervals for the photometric redshift estimates.  The
errors on the spectroscopic measurements are smaller than the plotting
symbols.  The six clusters with the largest redshift difference are
labeled.  \label{fig:zcomp}}
\end{figure}

We close this section with an assessment of the photometric redshift
estimates of V98 versus the spectroscopic measurements reported here.
Looking at the relevant data plotted in \mbox{Figure \ref{fig:zcomp}}
we see that in general the photometric redshifts were quite reliable.
In retrospect the 90\% confidence interval of $\Delta
z$\,=\,$^{+0.04}_{-0.07}$ for the 117 CCD-based estimates was somewhat
underestimated by V98, since only 70\% of the spectroscopic redshifts
actually fall within this error range.  Ignoring the six principal
outlyers labeled in \mbox{Figure \ref{fig:zcomp}}, the true 90\%
confidence interval is about $\Delta z$\,=\,$\pm 0.1$.

%
\section{Revised Cluster Catalog with Spectroscopic Redshifts and 
X-ray Luminosities}
\label{updatedsample}

We present the revised cluster catalog for the 160SD survey in Table
\ref{table:catalog}.  The most significant feature is the nearly
complete spectroscopic coverage.  We report spectroscopic redshifts
for 200 of the 201 (99.5\%) clusters in our sample.  With knowledge of
the cluster distances we can compute accurate X-ray luminosities.
Hence the fundamental parameters are now available to pursue detailed
investigations with this statistically complete, X-ray flux-limited
sample.



In Table \ref{table:catalog} we provide detailed information for each
of the 201 clusters discovered in the 160SD survey.  For completeness
we also list the 22 other objects meeting our original selection
criteria of significant X-ray extent.  Twenty-one of these, flagged
with ``F'' in columns (10), (11), and (13), are likely false
detections due to blends of unresolved X-ray point sources.  The other
source is only 4.4\arcmin\, away from the zero magnitude star Arcturus
and hence nearly impossible to confirm as a cluster.  

The object name and object number are given in columns (1) and (2).
Note the latter is the same identification used by V98.  The right
ascension and declination (J2000) for the centroid of the X-ray
emission are listed in columns (3) and (4).  Column (5) gives the
positional uncertainty in terms of the radius for the 90\% confidence
X-ray position error circle.  Columns (6) and (7) give the angular
core radius and its uncertainty based on a $\beta$-model fit with
$\beta=2/3$.  The total, unabsorbed X-ray flux in the 0.5--2.0 keV
energy band in the observer's restframe and its uncertainty are listed
in columns (8) and (9).

\begin{figure}
\plotone{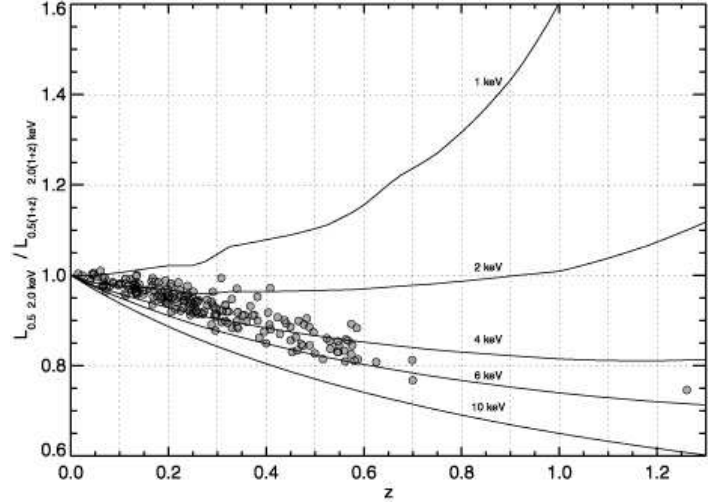}
\caption{{\em K}-corrections as a function of redshift and ICM
temperature in the \mbox{0.5 -- 2.0 keV} band.  The model source
spectrum is a Raymond-Smith plasma with a metallicity of 0.3 solar.
Data points indicate the sampling of the 160SD clusters.  The trend of
increasing cluster temperature with increasing redshift is the result
of the flux-limited nature of the survey.  \label{fig:kcorr}}
\end{figure}

Cluster X-ray luminosity in the \mbox{0.5--2.0 keV} energy band
is reported in column (10).  Luminosity in the object's rest frame
is defined by
\begin{equation}
L_{\rm X} ~=~  4\pi d_{\rm L}^{2}~f_{\rm X}~k_{\rm \,0.5-2.0}
\label{lx}
\end{equation}
where $f_{\rm X}$ is the total X-ray flux, $d_{\rm L}$ is the luminosity distance
\citep[e.g.,][]{Weinberg1972}, and $k_{\rm \,0.5-2.0}$ is
the \mbox{{\em K}-correction} in the \mbox{0.5 -- 2.0 keV} band.  The
latter transforms the {\em ROSAT\/} rest-frame luminosity into the
object's rest-frame luminosity.  The \mbox{{\em K}-correction} here is
given by

\begin{equation}
k_{\rm \,0.5-2.0} ~=~ \frac{\int_{0.5}^{2.0} f_{\rm E} d{\rm E}} 
                           {\int_{0.5(1+z)}^{2.0(1+z)} f_{\rm E} d{\rm E}}
\label{kcorr}
\end{equation}
where $f_{\rm E}$ is the differential flux (flux per unit energy) as a
function of energy and the integration limits are energy band edges in
keV.  The \mbox{{\em K}-corrections} for clusters, shown in
Figure~\ref{fig:kcorr}, were computed assuming a Raymond-Smith plasma
spectrum \citep{Raymond1977} with a metallicity of 0.3 solar and a gas
temperature for the intracluster medium (ICM) consistent with the
$L_{\rm X}$--$kT$ relation of \citet*{White1997},
\begin{equation}
kT = 2.76\,{\rm keV}~L_{\rm X,bol,44}^{0.33},
\label{lxkt}
\end{equation}
\noindent where $L_{\rm X,bol,44}$ is the bolometric X-ray luminosity
in units of \mbox{$10^{44}$ erg s$^{-1}$} with  $H_{0} = 50$ \hpone km s$^{-1}$ Mpc$^{-1}$

Heliocentric spectroscopic redshifts which entirely supersede the
photometric estimates of V98 are given in \mbox{column (11)}.  We
measured redshifts for 155 clusters while 45 come from the literature
and private communications as referenced in column (12).  In a few
cases the precision of literature-based redshifts reported by V98 have
been updated to reflect more robust measurements that are now available.

The final column (13) lists notes on individual clusters.  In
particular we indicate where our 160SD clusters are also members of
other X-ray selected samples.  Coincidences were found with the
BCS+eBCS \citep{Ebeling1998,Ebeling2000b}, BMW-HRI (L. Guzzo,
priv. comm.), BSHARC \citep{Romer2000}, EMSS \citep{Gioia1994}, NORAS
\citep{Boehringer2000}, RBS \citep{Schwope2000}, RDCS (P. Rosati,
priv. comm.), \mbox{REFLEX-I/II} (H. B\"{o}hringer, priv. comm.),
RIXOS \citep{Mason2000}, SSHARC \citep{Burke2003}, and
\mbox{WARPS-I/II} (\citealt{Perlman2002}; L. Jones, priv. comm.).  No
matches were found in the MACS (H. Ebeling, priv. comm.), the NEP
\citep{Mullis2001b, Gioia2003}, RASS1BS \citep{DeGrandi1999b} or the
SGP \citep{Cruddace2002,Cruddace2003} samples.

We also mark in column (13) the four X-ray overluminous elliptical
galaxies (OLEGs) described by \citet{Vikhlinin1999b} which are
potential ``fossil groups'' similar to those reported by others
(e.g., \citealt*{Jones2000b}, \citealt{Romer2000},
\citealt{Mullis2001b}).  Finally, we note nine instances where the
cluster redshift is within $\Delta$$z=0.015$ of the original target of
the {\em ROSAT\/} PSPC pointed observations (flagged with ``$z_{\rm
\,PSPC}$'').

\section{Discussion}
\label{discussion}

\begin{figure}
\plotone{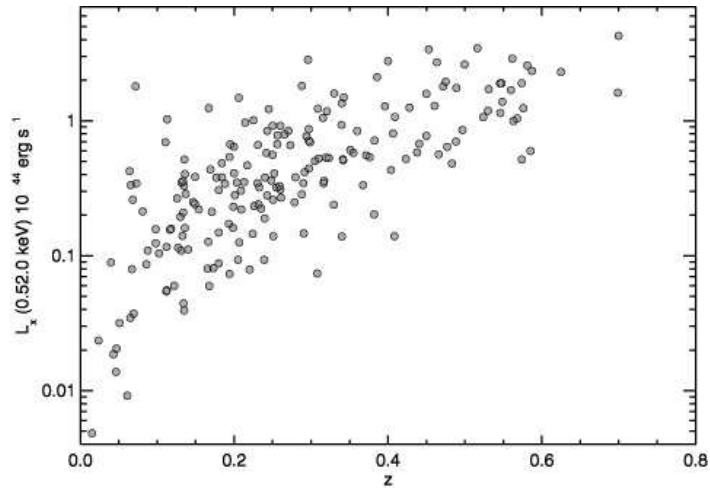}
\caption{X-ray luminosity and redshift distribution of the 160SD
cluster sample.  The median redshift is $z_{\rm median}=0.25$ and the
median X-ray luminosity is $L_{\rm X,median} = 4.2 \times 10^{43}$
\lxunits.  To preserve the readability of this plot the most distant
cluster (RX\,J0848.9+4452, $z=1.261$, $L_{\rm X}=2.0 \times 10^{44}$
\lxunits) is not shown.  
\label{fig:lxz}}
\end{figure}

We describe here the general properties of our cluster sample and
compare them to the results of other X-ray surveys.  In Figure
\ref{fig:lxz} we plot the luminosity-redshift distribution for 160SD
clusters.  The median redshift is $z_{\rm median}=0.25$ and the median
X-ray luminosity is $L_{\rm X,median} = 4.2 \times 10^{43}$ \hmtwo
\lxunits.  Note that the 160SD sample is the largest high-redshift
sample of X-ray-selected clusters published to date.  For example
there are 73 clusters at $z>0.3$ and 22 clusters at $z>0.5$.

As remarked by \citet{Vikhlinin2003} the {\em ROSAT\/} fluxes (and by
inference luminosities) used in our survey construction show very good
agreement with those measured in deep {\em Chandra\/} observations.
For the six 160SD clusters observed by both X-ray telescopes (as of early
2003), the fluxes differ by $\le$13\%, with no systematic offset, and
always within the statistical uncertainties.  Comparisons of fluxes
independently measured from {\em ROSAT\/} observations demonstrate
consistent results.  For example, using appropriate core radii and
redshifts, \citet{Romer2000} report no systematic offsets between
their measurements and ours for 11 clusters which are common to the
BSHARC and 160SD surveys.  Examining 16 clusters detected in the 20
{\em ROSAT\/} fields surveyed by both WARPS-I and 160SD,
\citet{Perlman2002} find a mean flux ratio of $f_{\rm 160SD}/f_{\rm
WARPS} = 0.98 \pm 0.26$.  We compared our fluxes to those of the
SSHARC survey \citep{Burke2003} and find for the nine shared clusters
the mean flux ratio is $f_{\rm 160SD}/f_{\rm SSHARC} = 0.78 \pm 0.22$.
This tendency for the SSHARC fluxes to be somewhat larger than those
of the 160SD and other surveys is also noted by \citet{Burke2003}.

The statistics concerning the membership of the 160SD clusters in
other samples are summarized in Table \ref{tab:overlap}. Six
independent cluster surveys, including the 160SD, have been extracted
from the same parent dataset, the {\em ROSAT\/} archive of pointed
PSPC observations.  Thus a significant amount of shared objects is
anticipated and is confirmed.  Given the large solid angle and deep
flux limit of our survey, the 201 clusters of 160SD comprise the
largest sample derived from the PSPC archive and encompass
approximately 30\%--60\% of the clusters from similar surveys.  By
number the largest overlap is 46 clusters shared with the RDCS,
whereas by percentage the largest coincidence is the 58\% of RIXOS
clusters.  Not surprisingly there are much fewer matches between the
160SD catalog and the surveys based on the RASS.  The 160
deg$^{2}$ sky coverage of our survey is a very small fraction of
the near all-sky coverage of the RASS.  Furthermore, these RASS-based
surveys are largely dominated by nearby ($z \la 0.3$) bright clusters,
which were often the target of PSPC observations and thus excluded by
design from the 160SD sample.  Finally, there are only minor
duplications between the {\em ROSAT\/} PSPC and the {\em ROSAT\/} HRI
or {\em Einstein\/} IPC pointings, hence the overlaps with the EMSS
and BMW-HRI samples are small.

Since the 160SD cluster sample was the first of the {\em ROSAT\/}
surveys to be published (V98), subsequent groups have had the
opportunity to make detailed comparisons to our sample (e.g., BSHARC:
\citealt{Romer2000}; SSHARC: \citealt{Burke2003}; WARPS-I:
\citealt{Perlman2002}).  In each case the authors have made a
field-by-field evaluation of their X-ray source detections and cluster
identifications versus ours for the {\em ROSAT\/} pointings common to
both surveys.  We highlight and further develop the key results using
the revised 160SD catalog.

There are 201 {\em ROSAT\/} PSPC fields processed by both the BSHARC
and 160SD surveys (\citealt{Romer2000}, see their \S\,7.5).  Note the
BSHARC survey adopted a wide and shallow strategy covering
approximately 178 deg$^{2}$ above a flux limit of $\sim$$3 \times
10^{-13}$ erg s$^{-1}$. In the shared area there are twenty-one 160SD
clusters which are sufficiently bright as to appear in the BSHARC
survey.  Thirteen of these X-rays sources are recovered in the BSHARC
sample, whereas eight are missing.  Six clusters (\#9, \#42, \#107,
\#182, \#195, and \#207) were not included because of a failure to
meet a filling factor criterion, and two (\#110 and \#201 ) were
missed because they were not detected as extended
(\citealt{Romer2000}; K. Romer, priv. comm.).  The BSHARC filling
factor diagnostic was used to reject blends or ``percolation
runaways''; see \citet{Perlman2002} for a discussion on the impact of
this filter on survey completeness.  For the thirteen sources
successfully detected by the BSHARC survey, we note only two
deviations in the optical identifications.  First, RX\,J0947.7+0741
(\#75) is identified as a QSO at $z=0.63$, but we have measured
concordant galaxy redshifts at $z=0.625$ thus suggesting that the
extended X-ray emission is the result of the ICM.  The QSO was in fact
identified as a separate point source by our X-ray analysis
software. Second as previously discussed in \S\,\ref{newzs},
RX\,J1524.6+0957 (\#170) is listed as a group at $z=0.078$ whereas we
conclude that a $z=0.516$ cluster is the dominant source.
\citet{Romer2000} did not identify any additional clusters within
160SD survey boundaries which are not already part of the 160SD
catalog.

Recently \citet[see latter portions of their \S\,3.1 and
\S\,4]{Burke2003} described the results of the SSHARC project and
examined the areas of the sky jointly surveyed by both the SSHARC and
160SD.  Note that both surveys probe to similar lower flux limits
($\sim 4~\times~10^{-14}$ \fxunits) but the SSHARC was limited to a
relatively small solid angle of 17.7 deg$^{2}$.  There are sixteen
160SD clusters in the overlap region.  Nine of these are identified as
clusters in the SSHARC sample whereas seven are missing.  One 160SD
cluster (\#215), which is also a part of the RDCS sample, was not
detected in the SSHARC survey.  The remaining six (\#39, \#43, \#75,
\#83, \#124, and \#214) were flagged as extended X-ray sources by
\citet{Burke2003}; however they concluded that their optical follow-up
does not suggest the existence of a cluster in any of these cases.
Nonetheless, we have measured multiple concordant redshifts for each
of these six clusters.  Also note clusters \#83 and \#124 are
confirmed members of the WARPS-II and RDCS samples, respectively.
\citet{Burke2003} point out one SSHARC cluster (RX\,J0505.3-2849,
$z=0.509$) which should formally meet our selection criteria but was
not included in our catalog because it was detected as two separate
sources. We manually inspected the processed {\em ROSAT\/} fields
during survey construction and suspected this double source was an
incorrect result of the X-ray detection algorithm. However, no other
similar cases were found in the entire survey, and we opted not to
modify the detection software for the sake of this single relatively
X-ray faint cluster.

The first phase of the WARPS survey \citep[and references
therein]{Perlman2002} is similar in sky coverage to the SSHARC.
WARPS-I covered 16.2 deg$^{2}$ down to a flux limit of
$6.5~\times~10^{-14}$ \ergps in total flux.  The on-going second
phase, WARPS-II, extends the survey area to about 73 deg$^{2}$
(H. Ebeling, priv. comm.).  \citet{Perlman2002} scrutinized the 20
{\em ROSAT\/} PSPC fields included in both the 160SD and WARPS-I
surveys.  They found that the 16 cluster identifications in the
overlap region are in complete agreement.  Furthermore, we note
excellent accord in the cluster redshifts except in one case.  For
RX\,J0210.4-3929 (\#25) we measured a redshift for the BCG of
$z=0.165$ and obtained a low signal-to-noise spectrum of a second
galaxy that is consistent with the BCG redshift.  In the WARPS-I
catalog \citep{Perlman2002} the redshift is quoted as $z=0.273$ but is
noted to be uncertain.  In light of our new evidence the WARPS team
re-examined their low-quality spectra for this cluster and find their
data are consistent with the 160SD redshift (E. Perlman, priv. comm.).

Finally, we emphasize a very important conclusion that can be drawn
from the careful work of the WARPS program.  Their survey is unique
amongst the complement of cluster searches based upon the {\em
ROSAT\/} PSPC archive of pointed observations in that they pursued
optical identifications for extended X-ray sources as well as for {\em
non}-extended sources.  The others, including the 160SD survey, only
examined X-ray emitters which were significantly extended.  The fact
that the WARPS team found no clusters missed by the 160SD, and in
particular, no cluster was missed because it was not resolved as
extended, underscores the completeness of our survey strategy.


%

\section{Summary}

We present the revised catalog of 201 galaxy clusters for the 160SD
survey featuring spectroscopic redshifts for 99.5\% of the members.
This sample includes 30\%--60\% of the clusters from similar {\em
ROSAT\/} cluster surveys, and is currently the largest high-redshift
sample of X-ray select clusters in the public domain.  We review the
X-ray criteria used to locate these galaxy systems, describe the
optical imaging and spectroscopy used to classify them, and compare
our results with similar studies. 

\acknowledgements It is a pleasure to thank Piero Rosati, Hans
B\"{o}hringer, Harald Ebeling, Kathy Romer, Eric Perlman, Stefano
Ettori and Axel Schwope for fruitful discussions.  We are grateful to
Harald Ebeling and the WARPS team, Piero Rosati and the RDCS team, and
John Huchra for sharing cluster redshifts prior to publication
(respectively 3, 1, and 2 entries referenced in \mbox{Table
\ref{table:catalog}}). Kathy Romer and the BSHARC team kindly shared
galaxy redshifts for the foreground group associated with
RX\,J1524.6+0957. We thank Piero Rosati and the RDCS team for
providing the unpublished RDCS sample, and Laurence Jones and the
WARPS team for providing access to the unpublished WARPS-II sample to
permit cross-correlations with the 160SD clusters.  Similarly we
appreciate Luigi Guzzo and the BMW-HRI team for sharing the target
list for their on-going survey for comparison with the 160SD\@.  Hans
B\"{o}hringer kindly provided the cross correlation of the unpublished
\mbox{REFLEX-I/II} samples and the 160SD clusters. The referee is
thanked for providing comments that helped clarify the presentation of
this work.  The support of the time allocation committees at UH and
ESO, and the expertise of the dedicated personnel at the Mauna Kea and
La Silla observatories are gratefully acknowledged.  The authors
recognize the religious and cultural significance of the summit of
Mauna Kea to the indigenous people of Hawai`i.  We deeply appreciate
the opportunity to pursue scientific research from this special place.

C.R.M. acknowledges partial financial support from NASA grant
NGT5-50175, the ESO Office for Science, and the ARCS Foundation. The
contributions of B.R.M., A.V., W.R.F., and C.J.F. were possible thanks
in part to NASA grant NAG5-9217 and contract NAS8-39073.  HQ was
partially supported by FONDAP Centro de Astrofisica. This research has
made use of the NASA/IPAC Extragalactic Database (NED) which is
operated by the Jet Propulsion Laboratory, California Institute of
Technology, under contract with the National Aeronautics and Space
Administration.

\bibliographystyle{apj}
\bibliography{myrefs}


%

%

%

%

%

%

%


%
\begin{deluxetable}{lll}
\tablewidth{0pt} 
\tabletypesize{\small}
\tablecaption{X-ray Selected Cluster Surveys\label{tab:xraysurveys}}
\tablehead{\colhead{Acronym\hfill} & \colhead{Name\hfill} & \colhead{Reference\hfill}}
\startdata
160SD    & 160 Square Degree {\em ROSAT\/} Cluster Survey                                    & \citealt{Vikhlinin1998}; this paper\\
BCS+eBCS & Brightest Cluster Sample + Extended Brightest Cluster Sample              & \citealt{Ebeling1998,Ebeling2000b}   \\
BMW-HRI	 & Brera Multiscale Wavelet {\em ROSAT\/} HRI Cluster Survey      & \citealt{Moretti2001,Panzera2003} \\
BSHARC   & Bright Serendipitous High-Redshift Archival {\em ROSAT\/} Cluster Survey  & \citealt{Romer2000}      \\
CIZA     & Clusters in the Zone of Avoidance Survey                                  & \citealt*{Ebeling2002}   \\
EMSS     & {\em Einstein} Extended Medium Sensitivity Survey                         & \citealt{Gioia1990a,Henry1992}\\
MACS	 & Massive Cluster Survey                                                    & \citealt*{Ebeling2001c}   \\
NEP      & {\em ROSAT\/} North Ecliptic Pole Survey                                  & \citealt{Henry2001,Mullis2001b}      \\
\enddata
\end{deluxetable}

%
\begin{deluxetable}{lcc}
\tablewidth{0in} 
\tablecaption{Optical Identifications of 160SD X-ray Sources\label{tab:identifications}}
\tablehead{\colhead{} & \colhead{V98\hfill} & \colhead{Revised Sample\hfill}\\
           \colhead{} & \colhead{\hfill} & \colhead{(this work)\hfill}}
\startdata
extended X-ray sources    & 223 & 223 \\
 \phantom{00}galaxy clusters        & \phantom{00}203 & \phantom{00}201 \\
 \phantom{00}probable false detections\tablenotemark{a} &  \phantom{002}19 &  \phantom{002}21 \\
 \phantom{00}no follow-up observations\tablenotemark{b} &  \phantom{0011}1 &  \phantom{0011}1 \\
\enddata
\tablenotetext{a}{likely blends of unresolved X-ray point sources}
\tablenotetext{b}{obscured by the glare of Arcturus}
\end{deluxetable}

%
\begin{deluxetable}{lccc}
\tablewidth{0in} 
\tablecaption{Redshifts for 160SD Clusters\label{tab:redshifts}}
\tablehead{\colhead{} & \colhead{V98\hfill} & \colhead{Newly Reported} & \colhead{Revised Sample\hfill}\\
           \colhead{} & \colhead{\hfill} & \colhead{(this work)\hfill} & \colhead{(this work)\hfill}}
\startdata
spectroscopic redshifts                      & \phantom{1}76 & 124           &200           \\
\phantom{00}measured by our group      & \phantom{00}\phantom{0}45 &\phantom{00}110           & \phantom{00}155           \\
 \phantom{00}literature \& private comm. &  \phantom{00}\phantom{1}31 &  \phantom{00}\phantom{0}14 &  \phantom{00}\phantom{1}45 \\
photometric estimates                       & 124           & $\cdots$      & $\cdots$      \\
clusters with no redshift                  & \phantom{12}3 & $\cdots$  & \phantom{12}1  \\
cluster sample size                         & 203           &    $\cdots$        & 201           \\
\enddata
\end{deluxetable}%


\end{document}